# eQTL Studies: from Bulk Tissues to Single Cells

Jingfei Zhang[a], Hongyu Zhao[b]

[a] Information Systems and Operations Management, Emory University
[b] Department of Biostatistics, Yale University

## Abstract

An expression quantitative trait locus (eQTL) is a chromosomal region where genetic variants are associated with the expression levels of certain genes that can be both nearby or distant. The identifications of eQTLs for different tissues, cell types, and contexts have led to better understanding of the dynamic regulations of gene expressions and implications of functional genes and variants for complex traits and diseases. Although most eQTL studies to date have been performed on data collected from bulk tissues, recent studies have demonstrated the importance of cell-type-specific and context-dependent gene regulations in biological processes and disease mechanisms. In this review, we discuss statistical methods that have been developed to enable the detections of cell-type-specific and context-dependent eQTLs from bulk tissues, purified cell types, and single cells. We also discuss the limitations of the current methods and future research opportunities.

## 1. Introduction

Recent years have seen great progress of identifying genomic regions associated with complex traits and diseases through genome wide association studies (GWAS), where tens of thousands of genomic regions have been associated thousands of traits, including many complex diseases (see the curated GWAS results at https://www.ebi.ac.uk/gwas/). One challenge of interpreting GWAS findings is that the majority of associated genetic variants, e.g. single nucleotide polymorphisms (SNPs), are in intergenic regions, making it difficult to infer functional genes and variants in these regions. Many efforts have been made to annotate the human genome through experimental (e.g. ENCODE project (CONSORTIUM 2012; CONSORTIUM et al. 2020); Roadmap Epigenomics Project (ROADMAP EPIGENOMICS et al. 2015); psychENCODE (ZHU et al. 2018)) and computational approaches (e.g. CADD (KIRCHER et al. 2014); GWAVA (RITCHIE et al. 2014); GenoCanyon (LU et al. 2015); GenoSkyline (LU et al. 2016); EIGEN (IONITA-LAZA et al. 2016); GenoSkyline-Plus (LU et al. 2017); STARR (LI et al. 2022)) to infer the functional roles of different SNPs and other variants, including eQTL studies (e.g. (CONSORTIUM 2020)), where the goal is to infer genetic variants affecting genetic regulation by associating genotypes with gene expression levels across a sample of individuals. Because eQTL studies measure expression levels of all the genes in the genome, they provide an unbiased view of the regulation of gene expression. Using results from eQTL studies in lymphoblastoid cell lines from HapMap samples, it was shown that SNPs associated with complex traits are significantly more likely to be eQTLs identified than minor-allele-frequency-matched SNPs (NICOLAE et al. 2010). In another study (KINDT et al. 2013) assessed enrichment and depletion of variants in different annotation classes, including genic regions, regulatory features, measures of conservation, and patterns of histone modifications. It was found that annotations associated with chromatin state together with eQTLs were the most enriched group. These early results stimulated many large community efforts to collect gene expression and genotype data for eQTL studies, and the accumulation of eQTL results parallels the great success of GWAS. If a SNP is associated with a complex trait, and is also associated with the expression level of a specific gene, this gene may be implicated as a possible candidate gene for the disease. A number of methods have been developed to formalize this idea for co-localization analysis that aims to find the SNPs that are associated with both expression and complex traits (HORMOZDIARI et al. 2016; WEN et al. 2016; GIAMBARTOLOMEI et al. 2018). Transcriptome wide association analysis methods have been developed to use eQTL data to predict the expression levels and associate the predicted

(imputed) expression levels with the observed complex traits (GAMAZON et al. 2015; GUSEV et al. 2016; HU et al. 2019). Mendelian randomization methods have also been proposed to investigate whether the expression trait is a causal factor for a complex trait of interest (RICHARDSON et al. 2020; YUAN et al. 2020; ZHOU et al. 2020; LIU et al. 2021b).

The most well-known eQTL study is the GTEx project where dozens of tissues from hundreds of individuals were analyzed to identify tissue specific eQTLs (CONSORTIUM 2020). The GTEx project has proved to be an extremely valuable resource for the research community. The version 8 of the GTEx analyzed 15,201 RNA-sequencing samples from 49 tissues of 838 postmortem donors. It was found that cis-eQTLs showed 1.46-fold enrichment in the GWAS catalogue (https://www.ebi.ac.uk/gwas/) where significant GWAS association results are collected. The cross tissue eQTL similarities were consistent with tissue relatedness, with testis, lymphoblastoid cell lines, whole blood, and liver distinct from other tissues, tissues from the brain region forming one cluster, and other organs being more similar to each other. BLUEPRINT collects genetic, epigenetic, and transcriptomic profiling in three immune cell types to investigate the contributions of different factors in gene expression (CHEN et al. 2016). eQTL catalogue is a resource developed by reprocessing data from dozens of studies with more than 30,000 samples in total, where summary statistics are available for many cell types and tissues (KERIMOV et al. 2021). The results from these studies and resources thus generated have demonstrated the values of eQTL information in inferring causal genes and variants at GWAS loci.

Most eQTL studies to date have been performed on bulk samples, where the estimated effect size of a SNP represents the average effects across different cell types, and the cell-type origin(origins) of the inferred eQTLs is(are) often unknown for a bulk sample consisting of distinct cell types. Despite some successes of using eQTL results to infer disease causing genes and variants, recent studies based on both modeling (YAO et al. 2020) and carefully chosen gene-trait pairs (CONNALLY et al. 2022) have shown that the known eQTLs, which are mostly derived from analysis of bulk tissues, only explain a very small proportion of the GWAS signals, where only a small proportion of the GWAS hits colocalize with eQTL SNPs. There is a growing evidence (as summarized below) that eQTL effects are often cell-type-specific and/or context-dependent, and many of the eQTLs uniquely identified through cell-type-specific and context-dependent analysis (either experimentally or computationally) colocalize with GWAS results (AGUIRRE-GAMBOA et al. 2020; DONOVAN et al. 2020; KIM-HELLMUTH et al. 2020; PATEL et

*al.* 2021), suggesting the importance of cell-type-specific and context-dependent eQTLs for interpreting and understanding GWAS signals. Therefore, there is a great need to identify these additional eQTLs missed from tissue-based analysis to expand the space of eQTLs and make more informed decisions on disease-causing genes and variants.

To facilitate the identifications of cell-type-specific and context-dependent eQTLs, statistical methods have been developed for both bulk samples through digital deconvolution analysis, and for single cell data, which offer finer cell type resolutions and can capture dynamic effects of eQTLs. In this review, we first discuss existing statistical methods that use bulk tissues and single cell data to identify cell-type-specific and context-dependent eQTLs. We then summarize results from empirical studies using bulk samples, purified cells, and single cell data. We conclude with the limitations of the existing computational methods and future methodological needs.

## 2. Analytical approaches for cell-type-specific and context-dependent eQTL inference

### 2.1. eQTL inference using bulk samples

Early eQTL studies collected gene expression data using microarrays, where gene expression levels need to be normalized to remove batch effects and the normalized data are analyzed to identify eQTLs. Consider a study with *N* subjects, *S* SNPs, and *G* genes. For the *n*th subject, let $y_{ng}$ denote the expression level of the *g*th gene, and $x_{ns}$ denote the genotype of the *s*th SNP. For a SNP with two alleles, say *A* and *a*, its three genotypes *AA*, *Aa*, and *aa* can be coded as 2, 1, and 0, respectively. The relationship between the observed gene expression level $y_{ng}$ and genotype $x_{ng}$ can be studied through the following regression model

$$y_{ng} = \beta_g + \beta_{gs} x_{ns} + \varepsilon_{ngs}, \qquad (1)$$

where $\beta_g$ is the intercept, $\beta_{gs}$ is the effect of the *s*th SNP on the expression of the *g*th gene, and $\varepsilon_{ngs}$ is the error term, often assumed to follow a normal distribution. A more comprehensive model may also include other covariates, such as age and sex. Testing the null hypothesis that the *s*th SNP does not have any effect on the expression level of the *g*th gene is equivalent to testing $\beta_{gs} = 0$. A typical eQTL study considers more than 20,000 genes and up to millions of SNPs. Because of the large number of SNPs to be tested, researchers often focus on cis-eQTLs for a given gene, which are SNPs in close physical proximity of the gene, say within one million base pairs of the candidate gene. In contrast, trans-sQTLs correspond to SNPs that are on different chromosomes or further away from the gene of interest on the same chromosome. Most eQTL findings have been for cis-eQTLs largely due to statistical power difference in detecting cis-eQTLs and trans-eQTLs. With a few hundred samples, which is the typical size of an eQTL study, there is limited power to do a genome wide association study required to identify trans-eQTLs, which often have smaller effect sizes than cis-eQTLs. Even for cis-eQTL analysis, there are often hundreds or thousands of SNPs that need to be considered, and multiple comparison adjustment needs to be done to appropriately control false positive findings. A number of computational tools have been developed and commonly used for eQTL analysis in bulk samples, such as MatrixEQTL (SHABALIN 2012) and FastQTL (ONGEN *et al.* 2016).

The regression setting in (1), where the errors are assumed to be Gaussian, is reasonable for microarray-based gene expression measurements. However, with gene expression data collected through RNA-sequencing, such as those from the GTEx project, the measured gene expression level is the total number of sequence reads mapped to a specific gene, which needs to be adjusted for total sequencing depth and other factors. These data may be better modeled by other distributions, e.g. negative binomial, while accounting for factors that may impact the observed sequencing reads. In this case, a generalized linear regression model may be more appropriate than (1) and may also have better statistical power, although it may be computationally more expensive.

For RNA-sequencing, there is added benefit of observing alternative splicing and allele-specific expression. In the case of allele-specific expression, consider the presence of a SNP in the transcribed region of a gene with two alleles *B* and *b*, and the simple scenario that all the sequence reads contain this SNP. For heterozygous individuals with genotype *Bb*, a sequence read covering this SNP may either have *B* or *b*. In one extreme case, all the sequence reads may only contain *B* but not *b*. Even in the absence of measured total gene expression levels for homozygous individuals with genotypes *BB* and *bb*, the imbalance between the mapped sequence reads having *B* and *b* suggests the presence of cis-eQTLs, either the SNP with alleles *B* and *b* itself or some SNP with perfect dependence with this SNP, that regulates the expression level of this gene. Statistical models have been proposed to explicitly incorporate this allelic specific expression to identify cis-eQTLs, including TReCASE (SUN 2012), RASQUAL (KUMASAKA *et al.* 2016), and mixQTL (LIANG *et al.* 2021). It was found that considering allelic specific expression could identify 20% to 100% more genes with eQTLs across 28 tissues in the GTEx project than only considering total expression levels using TreCASE (ZHABOTYNSKY et al. 2022), and the power gain of mixQTL was equivalent to a 29% increase in sample size for genes with sufficient allele-specific read coverage (LIANG et al. 2021).

The analysis of tissue level data also allows for the investigation of context-dependent eQTLs if the context can be well defined. For example, 369 sex-biased eQTLs were inferred through separate analyses of male and female GTEx samples (CONSORTIUM 2020), where the sex of an individual may be considered a context. Furthermore, 178 population-biased eQTLs were also implicated, where population origin may be considered another context. Other context-dependent effects can be considered by including an interaction term between the context variable and the SNP genotype in regression model (1).

## 2.2. Cell-type-specific eQTL inference using bulk samples

A number of studies have been published that use purified cells of different cell types to infer cell-type-specific eQTLs (ct-eQTLs). As the gene expression data in these samples are collected in the same manner as bulk tissue samples, the same statistical methods for bulk samples can be applied to infer ct-eQTLs from these data, although the sample size tends to be smaller and the measurement noises may be higher. In addition, the purified cell types may be contaminated with other cell types.

Without collecting data from purified cell types, Westra et al. (WESTRA et al. 2015) proposed to identify ct-eQTLs by investigating whether there is an interaction effect between the surrogate score for a cell type and candidate SNP's genotype on bulk gene expression levels from the collected samples. More formally, this model can be written as

$$y_{ng} = \beta_g + \beta_{gs}x_{ns} + \beta_{gm}m_n + \beta_{g,sm}(x_{ns} \times m_n) + \varepsilon_{ngs}, \qquad (2)$$

with two additional terms $\beta_{gm}m_n$ and $\beta_{g,sm}(x_{ns} \times m_n)$ compared to model (1), where $m_n$ is a proxy marker for the cell type of interest in the $n$th individual, $\beta_{gm}$ is the effect of the proxy marker on the expression level of the $g$th gene, and $\beta_{g,sm}$ is the interaction effect between genotype $x_{ns}$ and proxy marker $m$. A significant interaction effect, i.e. $\beta_{g,sm} \neq 0$, is interpreted as the cell-type-specific effect of SNP $s$ on expression of gene $g$. The same approach was used to study context-dependent eQTLs in (ZHERNAKOVA et al. 2017).

Instead of deriving cell-type-specific proxy markers or enrichment scores, the estimated cell type proportions can also be used as a proxy for a given cell type. Recent years have seen the developments of many methods to deconvolute bulk RNA-seq samples to infer proportions of different cell types and cell-type-specific expression levels (AVILA COBOS *et al.* 2020). For the *n*th subject, let $\pi_{nk}$ denote the estimated proportion of the *k*th cell type for this individual, where there is a total of *K* cell types. We can use the following regression model to detect cs-eQTLs for the *k*th cell type.

$$y_{ng} = \beta_g + \beta_{gs} x_{ns} + \beta_{gk} \pi_{nk} + \beta_{g,sk}(x_{ns} \times \pi_{nk}) + \varepsilon_{ngs}. \quad (3)$$

In this model, $\beta_{gk}$ is the cell type proportion effect from the *k*th cell type, and $\beta_{g,sk}$ is the interaction effect between the *s*th SNP and the proportion of *k*th cell type. A non-zero $\beta_{g,sk}$ suggests a cell-type-specific effect for the *s*th SNP.

The formulations in (2) and (3) consider one cell type at a time and ignore the contributions of possible cell-type-specific effects from other cell types, both in terms of proportions and the expression profiles, leading to a potential loss of information. Moreover, because models (2) and (3) only consider a tissue and cell type pair at a time, and may not attribute a non-zero $\beta_{g,sk}$ to the correct cell type. For example, consider the case where there are two cell types, where *k* = 1 or 2. If $\beta_{g,s1} > 0$, then $\beta_{g,s2} < 0$ due to the constraint that $\pi_{n1} + \pi_{n2} = 1$. Furthermore, the power differs across cell types, with a higher statistical power in detecting ct-eQTLs for more abundant cell types. A more comprehensive model that takes into account all cell types simultaneously can be formulated as

$$y_{ng} = \beta_g + \beta_{gs} x_{ns} + \sum_{k=1}^{K} \beta_{gk} \pi_{nk} + x_{ns}(\sum_{k=1}^{K} \beta_{g,sk} \times \pi_{nk}) + \varepsilon_{ng}, \quad (4)$$

subject to the constraint that $\sum_k \pi_{nk} = 1$. Correspondingly, the *s*th SNP is a ct-eQTL for the *k*th cell type if $\beta_{g,sk} \neq 0$. Another way to parametrize this model is in the form of

$$y_{ng} = \sum_{k=1}^{K}(\beta_{gk} + \beta_{g,sk} \times x_{nk})\pi_{nk} + \varepsilon_{ng}. \quad (5)$$

Note that the $\beta_{gk} + \beta_{g,sk} \times x_{nk}$ term in model (5) is essentially the cell-type-specific gene expression for the *k*th cell type in sample *n*, essentially the same model considered in Decon-eQTL (AGUIRRE-GAMBOA *et al.* 2020).

In practice, ct-eQTL analysis based on the above models often use transformed gene expression data instead of read counts and this may distort the association between the observed gene expression level with cell type compositions, leading to reduced power and inflated false positives. Recently, Little and colleagues proposed CSeQTL (Cell type-Specific eQTL) to jointly model total read counts and allele specific counts by a negative binomial (or Poisson) and a beta-binomial (or binomial) distribution with the consideration of covariates, cell type composition, and SNP genotype (LITTLE et al. 2022). CSeQTL also includes allele-specific expression to further increase the power to detect cell type-specific eQTLs. Empirical studies showed higher power of CSeQTL than linear model-based methods. In comparison, DeCAF is a linear model-based method that considers both total expression levels and allele-specific expression (KALITA AND GUSEV 2022).

Although the above approaches are intuitive, there are challenges to apply them to infer ct-eQTLs in practice. First, there are uncertainties in the estimated proxy markers and cell type proportions, and these need to be appropriately incorporated in the analysis. However, this issue has only recently been studied (CAI et al. 2022; XIE AND WANG 2022) and the impact of incorporating these uncertainty estimates in ct-eQTL inference needs to be studied. Second, although ct-eQTLs may be inferred for all cell types in principle, it would be relatively easier for more abundant cell types than for less abundant or rare cell types. Third, there has to be sufficient variations in cell type compositions across subjects to allow ct-eQTL inference. For example, in the extreme case that all the subjects have identical cell type proportions, the parameters in the above models are not identifiable. Fourth, the above formulation does not take into account similarity among some cell types, although methods have been proposed to consider cell lineage (YANKOVITZ et al. 2021).

**2.3. Cell-type-specific eQTLs from single cell data**

In addition to bulk data, single-cell data are increasingly used for ct-eQTL inference (JONKERS AND WIJMENGA 2017; VAN DER WIJST et al. 2018; LIU et al. 2021a; NEAVIN et al. 2021). Most published single-cell-based ct-eQTL analyses are performed through the analysis of pseudo-bulk RNA-seq data for different cell types, where the single cell data are first annotated to distinct cell types, and the cells annotated to the same cell types from a specific subject are combined together to derive cell-type-specific gene expression levels. eQTL methods for bulk samples can then be applied to detect ct-eQTLs. For example, (YAZAR et al. 2022) grouped cells of the same type for each individual and adjusted for covariate effects before using Spearman rank correlation analysis. However, the sample size is still much more limited for single-cell data compared to that of bulk samples and there are many ongoing efforts for single-cell-based genetic association analysis, e.g. the single-cell eQTLGen consortium (VAN DER WIJST *et al.* 2020) and the OneK1K cohort (YAZAR *et al.* 2022).

Instead of aggregating all the cells in a given individual, Nathan and colleagues (NATHAN *et al.* 2022) used Poisson mixed effects regression to model the effects of SNPs, cell states (which can be both discrete and continuous), batch structure, and other covariates (such as sex, age, genotype principal components and gene expression principal components, percentage of mitochondrial UMIs) on the observed gene expression level measured by UMI (unique molecular identifier) counts at the single cell level. The effect of a SNP is modeled as a fixed effect in the analysis. When the Poisson mixed effects model was compared with the computationally more expensive negative binomial mixed effects model, it was found that the Poisson model was adequate for the single cell data analyzed.

Although single cells can be grouped into pre-defined cell types for ct-eQTL analysis, the very high resolution at the single cell level offers the opportunity for more refined analysis, where the individual cells can be characterized by a vector of continuous contexts. For example, principal component analysis can be performed for the highly variable genes across all the cells based on normalized gene expression data, and the top principal components for a single cell may be taken as the cellular states for this cell. After the cellular states are defined for a single cell, the effects of a SNP on gene expression may be studied in the context of these cellular states to see whether the effects may vary depending on different states. Assuming we have *N* subjects, with $m_n$ cells collected from the *n*th subject, and there are a total of *C* different cellular contexts defined for each cell. Let the states of *i*th cell for the *n*th subject be denoted by a vector of contexts ($h_{ni1}$, $h_{ni2}$, …, $h_{niC}$) of dimension *C*. Cuomo and colleagues proposed a cellular regulatory map model, called CellRegMap, as

$$y_{ngi} = \beta_g + \beta_{gs}x_{ns} + \beta_{g,si}x_{ns} + u_{ng} + c_{ngi} + \varepsilon_{ngi}, \quad (6)$$

where $y_{ngi}$ represents the measured expression level of the *g*th gene in the *i*th cell of the *n*th subject, $x_{ns}$ is the genotype of the *s*th SNP of the *n*th individual, $\beta_g$ is the baseline expression level, $\beta_{gs}$ represents the persistent effect of the *s*th SNP across all the cells in different subjects, $\beta_{g,si}$ is the cell-specific effect on the *g*th expression level, $u_{ng}$ accounts for the fact that the $m_n$ cells are from the same subject, $c_{ngi}$ accounts for the cell context effects, and $\varepsilon_{ngi}$ is the error term. The CellRegMap model adopts an overall random effects model approach where $\beta_{g,si}$ ~$N(0, \sigma^2_{S \times C}\Sigma)$, $u_{ng}$~$N(0, \sigma^2_R \Sigma)$, $c_{ngi}$~$N(0, \sigma^2_C \Sigma)$, and $\varepsilon_{ngi}$~$N(0, \sigma^2_e)$. The matrix $\Sigma$ is defined by the cellular context vectors $\Sigma = HH^T$. CellRegMap uses a score test to investigate whether a SNP has context dependent effect on gene expression level with the null hypothesis $\beta_{g,si} = 0$. This model can also be used to test the main effect and estimate the allelic effects of single cells for each gene-SNP pairs based on the best linear unbiased predictor. In practice, it is important to define cellular contexts, and CellRegMap used MOFA (ARGELAGUET *et al.* 2018) to define cellular states, where latent factors are inferred from single cell data that explain variation in gene expression in the data. Because of the computational issues and the assumption of normal errors, the single cells were aggregated to meta cells because of the sparsity in single cell data in real data analysis. In addition, only specific gene-SNP pairs were considered due to statistical power concerns.

Strober and colleagues proposed a similar approach, called SURGE (Single-cell Unsupervised Regulation of Gene Expression), where a continuous representation of the cell contexts is learned through a probabilistic model with matrix factorization. The model has a form similar to that of CellRegMap as follows:

$$y_{ngi} = \beta_g + \beta_{gs} z_{ns} + \sum_{c=1}^{C} h_{nic} \beta_{gsc} z_{ns} + u_{ng} + \varepsilon_{ngi}, \quad (7)$$

where $y_{ngi}$ is standardized gene expression level for the *g*th gene in the *i*th cell of the *n*th subject and $z_{ns}$ is the standardized genotype for the *s*th SNP of the *n*th individual. The other parameters have the same meaning as the CellRegMap model, but the context vector $h_{nic}$ is latent and learned from the model instead of prespecified. SURGE has the following assumptions about the model parameters: $\beta_{gs} \sim N(0,1)$, $h_{nic} \sim N(0, \sigma_c^2)$, $\beta_{gsc} \sim N(0,1)$, $1/\sigma_c^2 \sim Gamm(\alpha_0, \beta_0)$, $u_{ng} \sim N(0, \psi_{gs}^2)$, $\varepsilon_{ngi} \sim N(0, \sigma_{gs}^2)$, $1/\psi_{gs}^2 \sim Gamm(\alpha_0, \beta_0)$, and $1/\sigma_{gs}^2 \sim Gamm(\alpha_0, \beta_0)$. SURGE approximates the posterior distribution of all latent variables using mean-field variational inference. Similar to CellRegMap, only eQTLs identified in previous studies were used for analysis due to statistical power concerns, and the reliance on the normal distribution assumption of the error terms limits its direct application to single cell data. As a result, single cells have to be aggregated to meta cells before the model is applied.

## 3. Empirical results on cell-type-specific and context-dependent eQTL analyses

In this section, we review the growing evidence of cell-type-specific and context-dependent eQTLs using bulk samples, purified cells, and single cells.

### 3.1. Tissue analysis

### 3.1.1. Whole blood

Among the first analysis of cs-eQTLs using bulk samples, (WESTRA *et al.* 2015) analyzed whole blood gene expression data of 5,683 individuals from seven cohorts to infer cell-type-specific cis-eQTLs. A total of 1,115 cis-eQTLs (8.5% of the significant cis-eQTLs from prior eQTL analysis for the whole tissue) were found to have significant interaction effects with neutrophil proxy. The results were replicated in six individual purified cell-type eQTL datasets. More importantly, the authors showed SNPs associated with Crohn's disease preferentially affect gene expression within neutrophils, demonstrating the insights gained from cell-type-specific eQTL analysis. (ZHERNAKOVA *et al.* 2017) performed eQTL and context-dependent eQTL analysis on RNA–seq data of peripheral blood from 2,116 unrelated individuals, identifying 23,060 genes with eQTLs, among which 2,743 (12%) showed context-dependent effects.

### 3.1.2. GTEx data

Cell-type-specific analysis was performed on the GTEx data in (KIM-HELLMUTH *et al.* 2020). The authors estimated cell type enrichment for seven cell types (adipocytes, epithelial cells, hepatocytes, keratinocytes, myocytes, neurons, and neutrophils) across 35 tissues. Between 43 pairs of tissues and cell types, they identified eQTLs specific to at least one cell type by testing for interaction effects between SNP and cell type enrichment on the observed expression levels. They found that these cell-type–interaction QTLs, called ieQTLs, are enriched for genes with tissue specific eQTLs and generally not shared across unrelated tissues. Furthermore, these ieQTLs are enriched for complex trait associations and had colocalization signals for hundreds of loci that were undetected in bulk tissue.

## 3.2. Cultured and purified cells

### 3.2.1. Brain cells

(AYGUN *et al.* 2021) used a cell-type-specific in vitro model system including 85 neural progenitors and 74 virally labeled and sorted neuronal progeny for eQTL analysis. They identified 2,079 and 872 eQTLs in progenitors and neurons, respectively, with 66% and 47% of these eQTLs not identified in fetal bulk brain eQTLs from a largely overlapping sample or in adult data from GTEx. These eQTLs had cell-type-specific colocalizations with GWAS hits for neuropsychiatric disorders and other brain-related traits.

Microglia in the brain play critical roles in immune defense and development, and are implicated in neurodegenerative disorders. (YOUNG *et al.* 2021) gathered gene expression profiles in primary microglia isolated from 141 patients undergoing neurosurgery. A total of 585 microglia eQTLs were identified. Through joint analysis with monocytes and IPSDMac, 855 microglia eQTLs were inferred, with 108 microglia specific, and 449 shared across three cell types. For colocalization with GWAS hits, there was an excess of colocalized microglial eQTLs for Alzheimer's disease, Parkinson's disease, and inflammatory bowel disease.

### 3.2.2. Melanocyte cultures

Because melanocytes give rise to melanoma but account for less than 5% of human skin biopsies, (ZHANG *et al.* 2018) performed eQTL analysis in primary melanocyte cultures from 106 newborn males to identify eQTLs in melanocytes. The identified melanocyte eQTLs differed considerably from those from the GTEx tissues, including skin. Novel risk genes for melanoma were implicated using the transcriptome wide association study based on this data set.

### 3.2.3. Immune cells

In the DICE project, 13 immune cell types were isolated from 106 leukapheresis samples of 91 healthy subjects (SCHMIEDEL *et al.* 2018). It was found that eQTLs are highly cell type specific, and sex has a major effect on gene expression. In the ImmuNexUT study, with samples from 79 healthy controls and 337 patients diagnosed with different immune-mediated diseases, (OTA *et al.* 2021) purified 28 immune cell types from these individuals with a total of 9,852 samples, and performed cell-type-specific eQTL analysis. They identified a median of 7,092 genes with eQTLs in each cell type, 2.2-fold more than that identified in the DICE study (SCHMIEDEL *et al.* 2018). They further identified eQTLs that were only present in patients.

## 3.3. Single cell analysis

### 3.3.1. PBMC

In a proof of concept study, (VAN DER WIJST *et al.* 2018) analyzed 25,000 single cell RNA-seq data from 45 donors. In total, they identified 379 unique cis-eQTLs involving 287 unique eGenes across six cell types. A total of 48 cis-eQTLs were only identified from cell-type-specific analysis. The authors also demonstrated the benefit of performing cell subtype analysis for cMonocytes and ncMonocytes.

In (OELEN *et al.* 2022), the authors exposed PBMC samples from 120 individuals to three pathogens and sequenced these samples in an unstimulated condition and after 3 hours and 24 hours in vitro stimulation for the three pathogens. They identified cell-type-specific eQTLs, with the number of such eQTLs correlated with the cell type abundance. Furthermore, the effects of eQTLs differed across pathogen stimulations, the strongest enrichment for GWAS signals was observed for eQTLs that were identified from stimulation experiments.

The investigators of the OneK1K cohort analyzed 1.27 million PBMC single cell RNA-seq data from 982 donors of Northern European ancestry and performed eQTL analyses on 14 immune cell types (YAZAR *et al.* 2022). A total of 26,597 cis-eQTLs were identified, with most having cell-type-specific effects. Dynamic effects were also observed based on pseudo-time trajectory for the B cell landscape. In addition to cis-eQTLs, 990 trans-eQTLs were identified, with most genes regulated by trans-eQTLs being specific for a single cell type, and none were ubiquitous across cell types. Co-localization analysis between eQTLs and GWAS signals suggested that 60% of colocalizing genes were detected upon activation and co-localization is very cell type-specific.

### 3.3.2. Induced pluripotent stem cells (iPSCs)

(NEAVIN *et al.* 2021) gathered single cell RNA-seq data from 64,018 fibroblasts of 79 donors and performed single cell eQTL analysis. For the six types of fibroblasts and four types of iPSCs, the majority of detected eQTLs in fibroblasts were specific to one cell type. Only 41% of the 45,503 eQTLs identified in the six fibroblast types were significant in GTEx,

Using 125 iPSC lines derived from 125 donors, (CUOMO *et al.* 2020) collected single cell gene expression data from 36,044 cells at four differentiation time points using full-length RNA-sequencing as well as the expression levels of selected cell surface markers through cell sorting. Substantial regulatory changes were observed with over 30% of eQTLs being specific to a single stage. Hundreds of eQTLs at the mesendo and defendo stages were new. This study also tested for associations between pseudo-time and the genetic effect size using a linear model, and identified 899 time-dynamic eQTLs.

### 3.3.3. T cells

(SOSKIC *et al.* 2022) analyzed 655,349 CD4+ T cells from 119 healthy donors, both for unstimulated cells and three time points after cell activation. Different numbers of genes showing eQTL effects were detected at different time points with hundreds of them only detected at specific cell states. Using pseudo-time trajectory information, 2,265 genes were found to have dynamic eQTL effects, representing about one third of the genes. Colocalized genes with GWAS signals were enriched in time-dependent eQTLs.

With 89 healthy donors, (SCHMIEDEL *et al.* 2022) performed eQTL analysis for more than one million activated CD4+ T cells classified into 19 distinct CD4+ T cell subsets. The effects of many eQTLs were strongly manifested only in certain cell types in an activation-dependent manner, and significant sex effects were also observed.

(NATHAN *et al.* 2022) performed single cell eQTL analysis using gene expression data from more than 500,000 unstimulated memory T cells from 259 Peruvian individuals. They found that the effects of one-third of cis-eQTLs were mediated by continuous multimodally defined cell states, with independent eQTLs at some loci having opposing cell-state relationships.

### 3.3.4. Brain cortex

In a recent study using single cell data from prefrontal cortex, temporal cortex and deep white matter from 192 individuals, a total of 7,607 genes were found to have eQTLs from eight cell types (BRYOIS *et al.* 2022). A majority of cell-type-specific eQTLs were replicated in tissue-level eQTLs for cortical tissue, with eQTLs for more abundant cell types more likely replicated in the tissue results. It was also found that the effect sizes estimated from tissue tend to be lower than those estimated from cell-type-specific analysis. As expected, the number of cis-eQTLs identified in a cell type was strongly correlated with the number of cells available for the corresponding cell type. The effect sizes were more similar for similar cell types with microglia being most different from other cell types. Co-localization analysis suggests that disease risk at a given GWAS locus is usually mediated by a single gene acting in a specific cell type.

### 3.3.5. Dopaminergic neuron differentiation

In (JERBER *et al.* 2021), the authors differentiated 215 human induced pluripotent stem cell (iPSC) lines to profile over 1 million cells across four conditions, including three differentiation stages (progenitor-like, young neurons and more mature neurons) and cells exposed to a chemical stressor. eQTL analysis was performed for 14 cell types, identifying 4,828 genes with eQTLs. Compared to eQTLs identified from GTEx brain tissues, this study identified 2,366 new eQTLs. As for colocalization analysis, 1,284 eQTLs were colocalized, with 597 being new, and 67% of these new colocalizations were associated with eQTLs detected in later differentiation stages or upon stimulation. A colocalization using aggregated data from different cell types yielded a much smaller number of colocalizations, suggesting the importance of considering cell type specificity.

## 4. Discussion

A main driving force for many eQTL studies in recent years is their potential to offer insights on the GWAS signals which mostly fall into non-coding regions of the human genome. Because of the importance of cell-type-specific and context-dependent eQTLs, there is a growing number of studies collecting and analyzing data to facilitate such analyses. Coupled with the rise of rich data that offer cell-type-specific and context-dependent gene regulation information, there is also a need for more statistically robust and computationally efficient methods for these analyses. In this paper, we have reviewed statistical methods that have been developed and applied to analyze different types of data for cell-type-specific and context-dependent eQTL inference, including bulk samples, purified cells, and single cells.

Despite these progresses, many issues remain to be resolved, especially in anticipation of the many population-level single cell data to be gathered in the near future that will involve tens of thousands of individuals and tens of millions of single cells across different tissues. For example, to fully respect the nature of the single cell data, the single data are more appropriately modeled as Poisson or negative binomial distribution. Yet, most published studies have adopted linear regression models and often aggregated multiple cells to address the sparsity of the observed single cell data. For bulk samples using RNA sequencing, efforts have been made to use allele-specific expression to improve statistical power for identifying eQTLs, and limited work has been done for single cell data to capitalize on allele-specific expression information. There are challenges in appropriately defining cell types and contexts for cell-type-specific and context-dependent eQTL analysis. The work of (CUOMO *et al.* 2022) and (STROBER *et al.* 2022) represent some initial efforts and more needs to be done to fully capture and utilize the cell state information for eQTL discoveries, including non-linear transcription programs (WANG AND ZHAO 2022). In addition, as more than one SNP may jointly affect expression levels (CONSORTIUM 2020; ABELL *et al.* 2022), methods that include joint effects are likely to be more powerful and can better characterize the relationship between gene expression levels and SNPs.

With many studies performed on bulk tissues, and the availability of a number of methods to use bulk tissue samples for cell-type-specific eQTL analysis, there is a need to effectively integrate the results from single cell data and bulk tissue data. Furthermore, as different tissues may share cells of similar cell types and states, there is also a need to better integrate results across tissues and studies (FLUTRE *et al.* 2013; URBUT *et al.* 2019).

As for all genetic studies, study design is important, such as the number of samples to be collected and, in the case of single cells, the number of cells per subject and the sequencing depth of each cell. A number of software and tools have been developed to facilitate this analysis under relatively simple statistical models for data analysis (MANDRIC *et al.* 2020; DONG *et al.* 2021; SCHMID *et al.* 2021). Further developments are needed to incorporate more comprehensive statistical models for analysis, and the consideration of other information, e.g. alleles-specific expression, in inferring eQTLs. Due to the limited statistical power, most studies to date have focused on cis-eQTLs. For example, fewer than 150 genes were found to be affected by trans-eQTLs in the GTEx project (CONSORTIUM 2020). There is a critical need to design statistical methods to identify trans-eQTLs, both for bulk tissues and for cell-type-specific and context-dependent effects.

We have focused on the inference of eQTLs in this paper. There are many downstream applications of eQTL analysis. For example, there is need for methods that perform co-localization analysis with cell-type-specific and context-dependent eQTL results without focusing on a specific set of SNPs through simple thresholding of statistical significance, while accounting for the linkage disequilibrium information across the SNPs in a region. Cell-type-specific and context-dependent eQTLs also offer the opportunity to improve the identifications of candidate genes through transcriptome wide association studies at the cell-type and context-dependent level, and the recently developed methods (SONG *et al.* 2023; OKAMOTO *et al.* 2023) for bulk samples can be extended for cell-type-specific and context-dependent analyses. There is also the need for Mendelian randomization methods to infer the causal relationship between transcript levels and complex traits and diseases (RICHARDSON *et al.* 2020; YUAN *et al.* 2020; ZHOU *et al.* 2020; LIU *et al.* 2021b). In this setting, the existing methods for bulk samples may not be adequate to deal with the count nature of the single cell data, data sparsity, and the need to integrate information from multiple data sources for more informed analysis and decision. In addition to eQTLs that affect gene expression levels, genetic variants can also affect gene expression variances and co-variances (HULSE AND CAI 2013; EK *et al.* 2018; SARKAR *et al.* 2019; MARDERSTEIN *et al.* 2021). Compared to eQTL analysis, relatively little has been done to identify such variants. Even less has been explored at the cell type level, where cell-type-specific co-expressions may be inferred using either bulk samples (SU *et al.* 2022b) or single cell data (SU *et al.* 2022a). A comprehensive catalog of eQTLs that have cell-type-specific and context-dependent effects on gene expression variance and co-variance will better characterize genetic regulation of expressions and interpret GWAS results.

Although gene expression has been the focus of cell-type-specific and context-dependent analysis, other data types are being increasingly collected for similar analysis, such as methylation, chromatin accessibility, and proteomics, based on single cell data, e.g. (WANG *et al.* 2022). The methods and tools developed for bulk and single cell RNA-seq data may also be applicable to other data types, such as recently published methylation data from the GTEx subjects (OLIVA *et al.* 2023) and an meQTL dataset derived from primary melanocytes of 106 individuals (ZHANG *et al.* 2021). More importantly, as different data types reflect different aspects of the same biological process, there is a need to integrate data from different modality to assess the genetic effects of SNPs on gene expression, methylation, chromatin accessibility, protein expression, and other molecular phenotypes. Such integrated analysis will likely yield more informative annotations of the SNPs to facilitate the interpretation of the GWAS results (HORMOZDIARI *et al.* 2018).

**Figure 1.** Illustration of eQTL analysis at different resolutions: **single cells**, **purified cells**, and **bulk samples**. Shown are data from three individuals with genotypes of AA, AG, and GG, respectively. There are two cell types making up the bulk samples, the oval shaped cells and the triangle shaped cells. For **single cell data**, we can observe expression level at the single cell level. For example, for the first individual with genotype AA, there are four oval shaped cells with expression level at 0.9, 1.1, 0.8, and 1.2, and two triangle shaped cells with expression level at 3.2 and 2.8, respectively. eQTL analysis can be performed for two cell types separately using single cells across these three individuals to correlate genotypes with observed single-cell level gene expression data. For data from **purified cells**, we observe aggregated gene expression levels for different cell types but without individual cell level measurements. The average expression level for the oval shaped cells is 1, 2, and 3, respectively, for the three individuals. For data from **bulk samples**, we can no longer distinguish contributions from two distinct cell types. The average expression level for the three individuals is 1.7, 2.0, and 1.7, respectively. For single cell data, not only we can study association between genotypes and cell-type-specific expressions, we can also correlate genotypes with cell type proportions. Through deconvolutions methods, the bulk samples may be deconvoluted to different cell types to allow cell-type-specific eQTL analysis with estimated cell type proportions from different individuals.

**Acknowledgments**

Zhang's research is supported by NSF DMS-2015190 and DMS-2210469. Zhao's research is supported in part by NIH R01 GM134005 and R56 AG074015.

# Figure 1

|  | Single Cells | Purified Cells | Bulk Samples |
|---|---|---|---|
| Individual 1 with genotype AA | 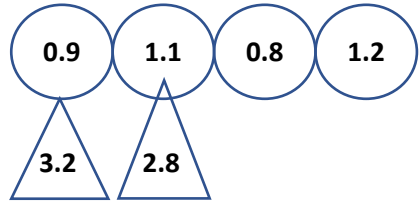 | 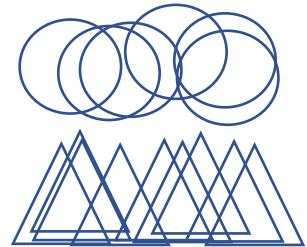 1.0 / 3.0 | 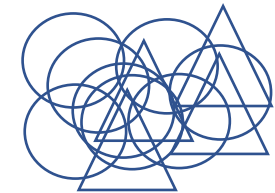 1.7 |
| Individual 2 with genotype AG | 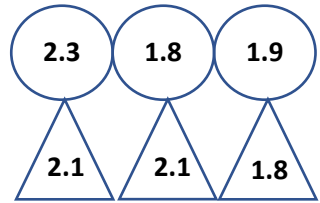 | 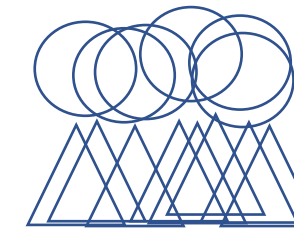 2.0 / 2.0 | 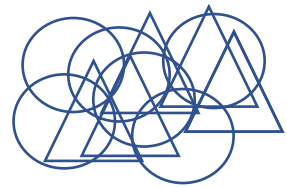 2.0 |
| Individual 3 with genotype GG | 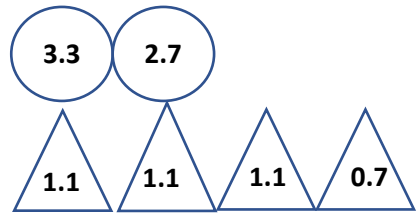 | 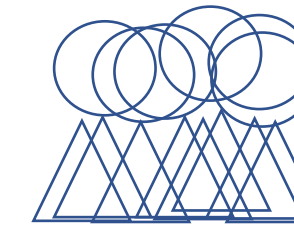 3.0 / 1.0 | 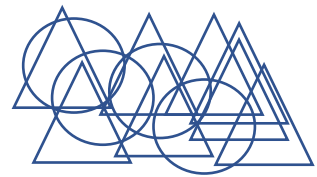 1.7 |